\documentstyle[sprocl,epsfig]{article}

\def\beq{\begin{equation}}
\def\eeq#1{\label{#1}\end{equation}}
\def\eeqn{\end{equation}}

\def\beqa{\begin{eqnarray}}
\def\eeqa#1{\label{#1}\end{eqnarray}}
\def\eeqan{\end{eqnarray}}

\let\bar=\overbar

\def\etal{{\it et al.}}

\def\VEV#1{\left\langle{ #1} \right\rangle}
\def\VEVS#1{\left[{ #1} \right]}

\def\Dslash{\not{\hbox{\kern-4pt $D$}}}
\def\dslash{\not{\hbox{\kern-2pt $\del$}}}

\def\ee{e^+e^-}

\def\msb{{\bar{\ssstyle M \kern -1pt S}}}

\def\Cpp{C\verb-++-}

\def\Title#1{\begin{center} {\Large #1 } \end{center}}
\def\Author#1{\begin{center}{ \sc #1} \end{center}}
\def\Address#1{\begin{center}{ \it #1} \end{center}}

\def\submit#1{\begin{center} #1 \end{center}}
\def\doeack{\footnote{Work supported by the Department of Energy,
                     contract DE--AC03--76SF00515.}}
\def\SLAC{Stanford Linear Accelerator Center\\
    Stanford University, Stanford, California 94309 USA}
\newcommand\pubblock{\rightline{\begin{tabular}{l} 
         SLAC-PUB-8289\\ October 1999 \end{tabular}}}
\newenvironment{Abstract}{\begin{quotation} \begin{center}
                       ABSTRACT
     \end{center}\bigskip  }{\end{quotation}}

\begin{document}
\begin{titlepage}
\pubblock

\vfill
\Title{Event Generators for Linear Collider Physics}
\vfill
\Author{Michael E. Peskin\doeack}
\Address{\SLAC}
\vfill
\begin{Abstract}
I review the array of event generators which have been written to provide
simulations of high-energy $\ee$ reactions.
\end{Abstract}
\vfill
\submit{presented at the International Workshop on Linear Colliders\\
          Sitges, Barcelona, Spain, 28 April -- 5 May 1999}

\vfill
\end{titlepage}
\def\thefootnote{\fnsymbol{footnote}}
\setcounter{footnote}{0}

\hbox to \hsize{\null}
\newpage
\setcounter{page}{1}

\title{EVENT GENERATORS FOR LINEAR COLLIDER PHYSICS}

\author{MICHAEL E. PESKIN}

\address{Stanford Linear Accelerator Center\\
    Stanford University, Stanford, California 94309 USA}

\maketitle\abstracts{I review the array of event generators which have
been written to provide simulations of high-energy  $\ee$ reactions.}

\section{Introduction}

In any simulation of an experimental analysis at an $\ee$ collider, we 
must begin with a sample of physics events to be analyzed. To produce
these, we need an event generator.  This program encodes our
knowledge of Standard Model background processes and our expectations for 
signal processes.  In this article, I will review the  current variety of 
event generators available for  simulations studies at future linear
colliders. 

There are a number of goals that an event generator might be expected to 
fulfill.  It should  realistically represent possible signal 
processes and Standard Model backgrounds.  It should take care of 
the superposition of QCD and fragmentation effects onto electroweak
cross sections.  It should give high accuracy, for precision studies.
And, it should have the flexibility to  include new reactions of 
arbitrary and exotic form.  These goals generally conflict with one  
another, or else are achieved only at the expense of a high level of 
complexity.  So we needed  different tools optimized for these various 
tasks.

In this report, I review the event generators which address these issues
that were presented at Sitges.
The major simulation programs described in this article are listed
in Table~\ref{tab:simtable}.  This table includes, for each program, a Web
address where download information and documentation can be  found.

\begin{table}[t]
\begin{center}
\caption{Event generators for $\ee$ linear collider physics }
\label{tab:simtable}
\medskip
\begin{tabbing}
 AA \= PANDORAXXX \=   \kill
 \>PYTHIA \> \verb+www.thep.lu.se/~torbjorn/Pythia.html+    \\
 \>HERWIG \> \verb+hepwww.rl.ac.uk/theory/seymour/herwig/+  \\
 \> CIRCE \> \verb+heplix.ikp.physik.tu-darmstadt.de/nlc/beam.html+ \\
 \> PHYSSIM \> \verb+www-jlc.kek.jp/subg/offl/physsim/+   \\ 
 \>PANDORA \> \verb+www.slac.stanford.edu/~mpeskin/LC/pandora.html+ \\
 \>ISAJET \> \verb+quark.phy.bnl.gov/~paige/+ \\ 
 \>SUSYGEN \> \verb+lyoinfo.in2p3.fr/susygen/susygen3.html+ \\
 \> EXCALIBUR \> \verb+home.cern.ch/charlton/excalibur/excalibur.html+ \\ 
\> KORALW \> \verb+hpjmiady.ifj.edu.pl/programs/node9.html+ \\ 
 \>WPHACT \>  \verb+www.to.infn.it/~ballestr+ \\ 
 \>KK   \>   \verb+home.cern.ch/~jadach/+ \\ 
 \>GRACE  \>  \verb+www-sc.kek.jp/minami/+ \\ 
 \>COMPHEP \>  \verb+theory.npi.msu.su/~comphep/+ 
\end{tabbing}
\end{center}
\end{table}

\section{Workhorses}

Among event generators, first place must be given to the general purpose
programs PYTHIA~\cite{Pythia} and HERWIG.\cite{Herwig,PH}  Both programs were 
originally written to  test ideas about QCD jet phenomena and hadronization.
But both have now evolved into general-purpose codes incorporating
all of the basic Standard Model processes in $\ee$ annihilation and a 
variety of nonstandard reactions.  

The most important aspect of PYTHIA and HERWIG is that they fully simulate
QCD final state state effects.  Given a system
 of two or more partons with large
invariant mass, these programs generate a QCD parton shower and then simulate
the hadronization of the final array of partons.  The shower algorithm is 
not exact at higher orders in $\alpha_s$ but does 
generate an approximately correct set of 
hard jets radiated from the original parton system. 
The hadronization step
is carried out by different algorithms in the the two programs, but, in 
both cases,  the description of hadronization has been tuned to fit the $\ee$
annihilation data.  These features imply that QCD final-state interactions 
have been included in a way that extrapolates correctly to high energy.

PYTHIA allows any parton-level generator to be included as a hard subprocess.
The generator must specific the color routing in the final state and the 
order in which parton showers are to be generated.
One caution with this 
approach is that order $\alpha_s$  corrections can raise or lower the overall
normalization of the cross section.  This effect cannot be included in the
hadronization but must be accounted for externally.

PYTHIA can run with a given initialization at a variety of $\ee$ center of 
mass energies. This allows the program to be linked to a generator of 
initial-state $e^-$ and $e^+$ energy distributions, such as CIRCE~\cite{OhlCC}
or PYBMS,\cite{BarklowPYBMS} to simulation the effect of beamstrahlung.
Initial state polarization is not included in the current version.  Final
state spin correlations are included for some but not all processes; notably,
spin correlations are included for the very important process $\ee\to W^+W^-$.

PYTHIA and HERWIG both include generators for the process $\gamma\gamma\to$
hadrons, including hard, soft, and `resolved' components.  The third of these
refers to processes in which partons in the photon undergo a hard scattering.
The relative magnitudes of these three components are not understood from 
theory, but it is important to understand this problem to compute the 
high-rate `minijet' background in which a $\gamma\gamma$ collision produces
a low-mass hadronic system.\cite{DRG,CBP}  New data on high energy 
$\gamma\gamma$ processes from LEP 2 and on $\gamma p$ processes from 
HERA---which contain much of the same physics--should allow systematic 
tuning of these generators.

\section{Polarization}

To introduce the next sections of this report, I must digress
on the subject of polarization.  Polarization has a central role in the LC
physics.  On one hand, because the LC will operate in the energy region well
above the $Z^0$ where it becomes manifest that left- and right-handed
have completely different quantum numbers, all standard and
non-standard cross sections will depend strongly on polarization. On the other 
hand, since it is difficult to measure polarization effects in the hadronic 
environment, polarization provides many new observables that cannot be 
studied at the LHC.  Thus it is important that both initial- and final-state
polarization be included properly in LC event generators.

How can polarization be included in physics simulations?  The traditional
approach is to generalize cross section formulae to include polarization 
asymmetries.  However, this rapidly becomes cumbersome.  Generators that 
take polarization seriously typically work at the amplitude level. 
 Even if one does not include polarization, it is 
useful to work with amplitudes in any complex Feynman diagram computation,
since if there are $N$ terms in the expression for the amplitude, there are
$N^2$ in the expression for the cross section.  Any 
 method that makes use
of amplitudes  to compute the cross section can be structured so that the 
polarization-dependence is available for free.

There are two common approaches for including polarization in the cross-section
formulae used in event generators.
The first approach is the {\em helicity-amplitude paradigm}.
In this approach, one computes amplitudes for transitions between states
of definite helicity.\cite{JandW}
  These amplitudes are then linked together to 
provide the complete amplitude for a process that turns the initial $\ee$
state into the final decay products.  Finally, the complete amplitude
is squared to provide the event weight.

The second approach is the {\em CALKUL paradigm}.  In this approach, one
concentrates on amplitudes with massless particles in the inital and final
states, the typical situation for $\ee$ annihilation when top quarks, $W$ 
bosons, and other heavy particles have decayed to their final products.
Then it is possible to compactly represent the amplitude for the 
whole process of production and 
decay
in terms of {\em spinor products},\cite{KS,MP}
\beq
    \VEV{12} = \bar u_L(p_1) v_R(p_2) \ ,\qquad 
 \VEVS{12} = \bar u_R(p_1) v_L(p_2) \ .
\eeq{spinorp}
The spinor products can in turn be computed directly for the
set of initial and final 
massless four-vectors in a given event.

It is important to note that there are no important polarization effects
associated with hadronization, except that the $\tau$ decay depends strongly
on $\tau$ polarization.  This effect should be accounted explicitly by 
decaying $\tau$'s through the simulation program TAUOLA.\cite{Tauola}

The use of helicity amplitudes to systematically describe LC physics was
pioneered by the JLC group, using the programs HELAS,\cite{HELAS}
 for automatic 
Feynman diagram computation, and 
BASES/SPRING,\cite{Kawabata} to provide weight-1 events. 
 The current version of their
package is PHYSSIM in Table~\ref{tab:simtable}.

The need to build up systematically
the full complexity of LC reactions---including 
beamstrahlung, initial-state radiation, initial- and final-state polarization
effects, and hadronization---has led the authors of almost all the 
simulation programs to embrace object-oriented programming for their 
future versions.  One relatively simple generator, pandora, 
already segregates the beam and $\ee$ reaction information into 
separate \Cpp\ classes which interact through a simple interface.
Further details can be found in ref. \cite{mypandora}.

\section{Supersymmetry}

The next few sections of this report will discuss generators devoted to 
specific problems of LC physics.  The first of these is the coherent 
representation of supersymmetry processes.  

The full set of processes in $\ee$ annnihilation to two supersymmetric 
particles  is now available in three
different  programs, the supersymmetric extension of PYTHIA,\cite{SPYTH}
the latest release of ISAJET,\cite{Paige} and the SUSYGEN program written
for LEP 2 studies.\cite{Ghodbane}  ISAJET was the first to include
polarization-dependent cross sections.  The new version also correctly 
includes the matrix elements for  3-body decays. SUSYGEN gives a complete
treatment of initial and final polarization effects using the 
helicity-amplitude paradigm and even allows for nonzero phases in the 
$A$ and gaugino mass parameters.  ISAJET and SUSYGEN explicitly include
parametrizations of beamstrahlung.  All three programs allow input of
a general set of supersymmetry parameters. Given the model-independent 
character of LC measurements, this is an important feature.
 PYTHIA and ISAJET also include facilities that
compute the supersymmetry spectrum from an underlying model; SUSYGEN
includes an interface to spectrum calculations with SUSPECT.\cite{Djouadi}

\section{Precision Standard Model}

For calculation of Standard Model background processes at a LC, it is not 
sufficient to consider $\ee$ annihilation to on-shell 2-body final states.
Backgrounds to new physics typically come from higher-order corrections
in which additional fermions are produced or from $\ee\to W^+W^-$ processes
in which one $W$ boson fluctuations far off the mass shell.  In fact, these
reactions are not distinct and one must include all $\ee\to 4$ fermion 
Feynman diagrams in order to obtain a gauge-invariant result. 

This is already an issue at LEP 2 and a very serious effort has been
made to provide 4-fermion event generators for the LEP 2 experiments.
The status of generators for 4-fermion and $W$ pair physics has recently
been described by Bardin, \etal\cite{WWCERN}  These programs typically use
the CALKUL paradigm.  Though their implemetations are slightly different, 
they agree excellently among themselves and with the LEP 2 for configurations
of four fermions all at large relative momenta.  For brevity, 
I have included only three of 
these programs, EXCALIBUR,\cite{Excalib} KORALW,\cite{koralw}, and
WPHACT,\cite{wphact} in Table~\ref{tab:simtable}.

Two unresolved conceptual problems in the simulation of 4-fermion processes are
the treatment of the $W$ width for an off-shell $W$ and the correct inclusion
of transverse momentum for almost-collinear initial state radiation.  In 
both cases, there is no simple prescription which is gauge-invariant.  This
leads to discrepancies among the various programs in certain specific kinematic
regions.  For example, for the process $\ee\to e^+ \nu d \bar u$ (very low
mass single $W^*$ production), the various generators give 10\% differences
in the predicted cross section when  $m(d\bar u)$ is as small as a few GeV.

Additional challenges can be found in  higher-order processes.  For the 
study of the standard process $\ee\to t \bar t$ and also for many searches,
one needs an event generator for $\ee\to 6$ fermions.  Accomando, 
Ballestrero, and Pizzio\cite{ABP} have computed the relevant cross sections
in a suitable form and are preparing a new generator, SIXPHACT.  Alternatively,
methods are now available to allow one to perform the computation 
automatically; I will describe these in the next section.

At the same time, it is necessary to reach for higher accuracy in the 
simulation of 2-fermion final states.  KORALZ, by Jadach, Ward, 
and Was,\cite{koralz} achieved an accuracy of 0.1\% in the calculation of 
the small-angle Bhabha scattering cross section, and this was essential
for the precision cross section normalization at LEP 1.  At higher
energies, it is necessary to treat multiple photon emissions coherently, 
and the precision calculations must be extended to larger forward angles.
Jadach and collaborators have just released a new program KK which
addresses these issues.

\section{Do it yourself}

Eventually, workers in LC physics will have a need for event generators for
processes that have not been included in the standard programs.  The 
traditional recourse in this case has been to find a friendly theorist with
time on his hands.  Today, however, amother course is made available by 
the GRACE\cite{grace} and COMPHEP~\cite{comphep} programs.  These encode
the Feynman rules of the standard model and certain extensions, and allow
encoding of arbitrary additional Lagrangian couplings, and then automatically
generate the numerical sum of Feynman diagrams.  The authors of  GRACE
have made available a supersymmetric extension which already includes all
239 $\ee\to 3$-body supersymmetry processes, and all 1424 
  $\ee\to 3$-body processes.  A typical process might involve the summation
of 100 tree diagrams.  The calculations are done at the amplitude level,
allowing the event generation to include full spin correlations using the 
helicity-amplitude paradigm.  These programs can also compute one-loop
corrections by evaluating diagrams in terms of the standard set of one-loop
integrals defined by Passarino and Veltman.\cite{PV}  A more complete
discussion of these systems can be found in Perret-Gallix's contribution 
to this conference.\cite{PG}

\section{Conclusions}

In this report, I have tried to summarize the array of programs that are now
available to perform  event generation for  LC physics.  These range from 
the general-purpose generators PYTHIA and HERWIG, to specific tools for 
supersymmetry and multi-fermion simulations, to tools for automatic 
generation of events for arbitrary physics processes.  For the future,
we expect to see trends toward object-oriented and modular programs, toward
detailed high-accuracy computation of standard background processes, and toward
further automation of complex calculations.  We are well on the way to the 
level of accuracy and generality that will be needed for the LC physics 
program.

\section*{Acknowledgments}
I am grateful to the authors of the programs described above
 for their help in organizing this 
review.  This 
work was supported by the US Department of Energy under contract
DE--AC03--76SF00515.

\end{document}